\documentclass[twocolumn,showpacs,preprintnumbers,amsmath,amssymb]{revtex4}
\usepackage{dcolumn}
\usepackage{bm}
\usepackage{graphicx}

\usepackage{hyperref}
\hypersetup{colorlinks,
linkcolor=blue,
filecolor=green,
urlcolor=blue,
citecolor=blue}

\def\bk{{\bf k}}
\def\br{{\bf r}}
\def\bx{{\bf x}}
\def\bE{{\bf E}}
\def\bR{{\bf R}}
\def\l{\lambda}
\def\Eklr{\mathbf{E}(\bk \lambda ;\br )}
\def\akl{a_{\bk \lambda}}
\def\akld{a_{\bk \lambda}^\dagger}

\begin{document}

\title{van der Waals interaction energy between two atoms moving with uniform acceleration}

\author{Antonio Noto and Roberto Passante}
\affiliation{Dipartimento di Fisica e Chimica dell'Universit\`{a} degli Studi di Palermo and CNISM, Via Archirafi 36, I-90123 Palermo, Italy}

\email{roberto.passante@unipa.it}


\begin{abstract}
We consider the interatomic van der Waals interaction energy between two neutral ground-state atoms moving in the vacuum space with the same uniform acceleration. We assume the acceleration orthogonal to their separation, so that their mutual distance remains constant.
Using a model for the van der Waals dispersion interaction based on the interaction between the instantaneous atomic dipole moments, which are induced and correlated by the zero-point field fluctuations, we evaluate the interaction energy between the two accelerating atoms in terms of quantities expressed in the laboratory reference frame. We find that the dependence of the van der Waals interaction between the atoms from the distance is different with respect to the case of atoms at rest, and the relation of our results with the Unruh effect is discussed. We show that in the near zone a new term proportional to $R^{-5}$ adds to the usual $R^{-6}$ behavior, and in the far zone a term proportional to $R^{-6}$ adds to the usual $R^{-7}$ behavior, making the interaction of a longer range. We also find that the interaction energy is time-dependent, and the physical meaning of this result is discussed. In particular, we find acceleration-dependent corrections to the $R^{-7}$ (far zone) and $R^{-6}$ (near zone) proportional to $a^2t^2/c^2$; this suggests that significant changes to the van der Waals interaction between the atoms could be obtained if sufficiently long times are taken, without necessity of the extremely high accelerations required by other known manifestations of the Unruh effect.
\end{abstract}

\maketitle

\section{\label{sec:1}Introduction}

van der Waals and Casimir-Polder forces are long-range interactions between atoms or between atoms and neutral macroscopic objects, respectively, due to fluctuations of the quantum electromagnetic field \cite{CP48,DMRR11}. These interactions have not a classical analogue, and they can be equivalently attributed to zero-point fluctuations of the quantum electromagnetic field or to fluctuations of currents in the microscopic and/or macroscopic objects considered \cite{Milonni07,Buhmann12}. They have been recently considered also in dynamical (time-dependent) situations \cite{MVP10}. Although these interactions have a very small strength, they have been measured in several physical situations \cite{BP99,BCMZS10,BVCLB13,OWAPSC07}.

One important aspect is related to the change of both van der Waals and Casimir-Polder interactions due to the motion of the objects considered. A uniformly accelerated motion is particularly relevant, due to the so-called Unruh effect: the Unruh effect predicts that a uniformly accelerated observer perceives vacuum fluctuations as a thermal field with temperature $T=\hbar a/(2\pi c k_B)$, $a$ being the observer's acceleration \cite{Unruh76,CHM08}. The Unruh effect has not been observed yet, although many proposals for its measurement have been presented in the literature (see \cite{CHM08} and references therein), for example detecting spin depolarization of accelerated electron \cite{BL83}, or accelerating particles by ultraintense laser pulses \cite{CT99,SCH06} or laser filaments \cite{US12,BCCGORRSF10}.
These considerations give important motivations for investigating also other observable physical effects that can be affected by an accelerated motion, and that could give evidence of the Unruh effect and in general quantum-electrodynamical effects related to non-inertial motion. Effect of a uniform acceleration on Lamb shift \cite{AM95,Passante98}, atom-wall Casimir-Polder interactions and related phenomena \cite{RS09,ZY10,ZYL06} has been investigated in the literature; it was shown that extremely high accelerations ($\sim 10^{22} {\rm m/s^2}$) are necessary in order to make observable the Unruh effect in the Lamb shift and atom-wall interactions.

van der Waals dispersion forces between two neutral atoms in the vacuum are related to fluctuations of the zero-point electromagnetic field, and thus they could be a good candidate for detecting an accelerated motion of the atoms and the Unruh effect. In this paper we will consider the effect of the acceleration on the dispersion interaction between two atoms, and we will show that new phenomena are present in this case, namely a change of the distance-dependence of the interaction energy and its explicit time-dependence. Using a simple model, we had already obtained some hints on the effect of the Unruh effect on the dispersion force on accelerating atoms, exploiting the relation between acceleration and temperature given by the Unruh effect \cite{MP10}.

As mentioned, in this paper we investigate the effect of a uniform acceleration on the van der Waals interaction energy between two ground-state atoms moving in the vacuum space with a uniform acceleration. In particular, we are interested to investigate whether the (uniform) acceleration of the atoms yields a qualitative change of the force properties.
We consider two atoms/molecules A and B moving, in the laboratory system, with the same uniform acceleration $a$ in the $x$ direction. They move along the same direction perpendicular to their distance, so that their separation is constant. In order to obtain their van der Waals interaction, both in the near zone and in the far zone (Casimir-Polder regime), and in particular how this interaction is affected by their acceleration, we use the following physical model: the interaction energy arises from the dipolar interaction between the (instantaneous) oscillating dipole moments of the atoms, induced and correlated by zero-point electromagnetic field fluctuations.
In this model the dipolar fields are classical fields, and the quantum properties of the radiation are included in the spatial correlations of the electric field associated to vacuum fluctuations.
This model has been used and proved valid for atoms at rest \cite{PT93,PPR03}, and it has been used also for three-body dispersion forces  \cite{CP97}, when boundary conditions are present \cite{SPR06} or in the presence of external radiation \cite{Salam09}. In the present case we need to generalize this model to the case of accelerating atoms, expressing the field generated by the atomic dipole moments in the accelerated reference frame. An advantage of our method is that, even if the interaction energy is calculated for the accelerating atoms in their co-moving frame
(the system in which the atoms are instantaneously at rest), all physical quantities relative to the atoms are given in terms of their known values in the laboratory frame.
We obtain an explicit expression of this interaction energy and discuss in detail the near- and far-zone limits, showing that main effects of the accelerated motion of the atoms are a change of the distance-dependence of the van der Waals interaction, with respect to the usual case of inertial atoms, and an explicit time-dependence of the interaction energy.
Our results show that in the near zone a new term proportional to $R^{-5}$ adds to the usual $R^{-6}$ behavior, and in the far zone a term proportional to $R^{-6}$ adds to the usual $R^{-7}$ behavior, making the interaction of a longer range. We also find that the interaction energy has an explicit time dependence. In particular, we show that acceleration-dependent corrections to the $R^{-7}$ (far zone) and $R^{-6}$ (near zone) terms, proportional to $a^2t^2/c^2$, are present. This suggests that significant changes to the interaction between the two atoms could be obtained if sufficiently long times are considered, even for reasonable values of the acceleration, contrarily to other known manifestations of the Unruh effect, such as Lamb shift and atom-wall interaction for accelerated atoms, which require extremely high accelerations. Both these effects could be relevant for the observation of the Unruh effect or other acceleration-dependent effects in quantum electrodynamics. Although Lamb shift and atom-wall interactions have the same physical origin of the atom-atom interaction, the van der Waals interaction seems more sensitive to the acceleration because of the time-dependence of the interaction in this case. This time dependence follows from the effective interaction distance introduced in Section \ref{sec:2} which grows with time for the atom-atom interaction, while it is constant in the case of an atom accelerating parallel to an infinite wall (such a concept has not meaning for the Lamb shift).

This paper is organized as follows. In Section \ref{sec:2} we introduce our physical model for the van der Waals interaction energy for the accelerating atoms, based on the method of correlated induced dipole moments, and evaluate the dipole fields of the accelerating atoms; we also introduce the important concept of {\it effective interaction distance}. In Section \ref{sec:3}, after appropriate Lorentz transformation to the co-moving reference frame, we use these results in order to evaluate the interaction energy between the two atoms in the accelerated frame, in terms of physical quantities calculated in the laboratory frame. Finally, Section \ref{sec:4} is devoted to the discussion of our results and some concluding remarks.

\section{\label{sec:2}The model for the van der Waals interaction in the accelerated frame}

In this Section we generalize the method of induced dipole moments, originally introduced for the calculations of the van der Waals interaction between atoms at rest \cite{PT93,PPR03}, to the case of atoms in accelerated motion. In this model, the interatomic interaction energy originates from the interaction of the instantaneous dipole moments of the two atoms. These dipoles are induced and correlated by the spatially correlated zero-point fluctuations of the quantum electromagnetic field. In this model the quantum nature of the dispersion interaction enters in the correlation function of zero-point fluctuations of the electric field, while the dipole fields are treated classically.

$\tilde{E}_i(\bk \l ,\br,  t)$ indicates the $(\bk \l )$ Fourier component ($\l =1,2$ is the polarization index) of the electric field at position $\br$, generated by atom A whose position is $\bR_A$. This field, evaluated in the moving reference frame where atom B is instantaneously at rest (i.e. the co-moving frame), depends on the instantaneous (fluctuating) dipole moment of atom A in the laboratory reference frame at the retarded time $t_r=t-\rho (t_r)/c$. $\rho (t_r)$ is an \emph{effective interaction distance} given by the distance traveled by a light signal from its emission by atom A at time $t_r$ to the time $t$ when it is received by atom B. We shall evaluate this distance for our specific case of uniformly accelerated atoms at the end of this Section. In this model, the atoms are assumed as having instantaneous oscillating dipole moments and their van der Waals interaction arises from the interaction between the field emitted by the fluctuating dipole of one atom with the dipole moment induced on the second atom. This field can be expressed as (summation over repeated index is understood),

\begin{equation}
\tilde{E}_i(\bk \l ,\bR_B,  t)=-\mu_j^A\tilde{V}'_{ij}(k,\bR,t_r) ,
\label{eq:1a}
\end{equation}
where $\bR_B$ is the position of atom B, $\bR = \bR_B-\bR_A$ and $\mu_j^A$ is the dipole moment of atom A. $\tilde{V}'_{ij}(k,\bR,t_r)$ is a tensor potential that will be obtained in the next Section. From now onwards, a \emph{tilde} indicates that the corresponding quantity is evaluated in the co-moving reference frame, where the atoms are instantaneously at rest. In the co-moving frame, the interaction of this field with the induced dipole moment of atom B is given by
\begin{equation}\label{eq:1b}
- \tilde{\mu}_i^B \tilde{E}_i(\bk \l ,\bR_B,  t)
=\tilde{\mu}_i^B\mu_j^A\tilde{V}'_{ij}(k,\bR ,t_r) ,
\end{equation}
where $\tilde{\mu}_i^B$ is the dipole moment of atom B in the accelerated frame. Summation over $(\bk \l )$ yields the interaction energy.

The Fourier $(\bk \lambda )$ component $\Eklr$ of the electric field, given by
\begin{eqnarray}
E_j(\bk \lambda ;\br)&=&i\left( \frac {2\pi \hbar ck}V \right)^{1/2} \Big( \hat{e}_j(\bk \lambda ) \akl e^{i\bk \cdot \br} \nonumber \\
&-&\hat{e}_j^\star(\bk \lambda )
\akld e^{-i\bk \cdot \br} \Big)
\label{eq:1}
\end{eqnarray}
($\hat{e}_j(\bk \lambda )$ is the polarization unit vector), induces a dipole moment in the atom at position $\br$ given by
\begin{equation}
\mu^{ind}(\bk \lambda ;\br) =\alpha (k) \Eklr ,
\label{eq:2}
\end{equation}
where we are assuming an isotropic atom with dynamic polarizability $\alpha (k)$. The instantaneous dipole moment induced in one atom, let us say atom A, generates an electric field that then interacts with the other atom (B). This electric field is the the field generated by atom A with position $\bR_A$ at the retarded time $t_r = t -\rho (t_r)/c$, evaluated at the position of atom B.
Because in our case both atoms are accelerating. we need the expression of the electric field generated by an oscillating dipole in motion. This expression, as well as that of the magnetic field, is known and it is usually separated in the two components $\bE^{(pol)}$ and $\bE^{(Roe)}$, called the polarization and R\"{o}ntgen components, respectively. Because we are interested in the interaction between the two accelerating atoms in their co-moving system, that is a locally inertial frame, the electric field must be Lorentz-transformed to the co-moving system: thus, electric and magnetic fields are both necessary, because Lorentz transformations mix electric and magnetic fields. In the laboratory frame, these fields, for a dipole moving along an arbitrary trajectory $\bx (t)$, are given in Ref. \cite{PT01,Thirunamachandran06} in terms of the retarded time $t_r=t-r/c$. We use the general expressions in \cite{PT01,Thirunamachandran06} for the polarization and R\"{o}ntgen components of the electric and magnetic fields in our case of a uniformly accelerated trajectory along $x$ given by \cite{Rindler91}
\begin{eqnarray}\label{eq:8}
x(t) &=& \frac{c^2}{a}\left( \sqrt{\frac{a^2 t^2}{c^2}+1} -1 \right), \nonumber \\
x(\tau ) &=& \frac{c^2}a \left( \cosh \frac{a \tau}c -1 \right) ,
\end{eqnarray}
where time $t$ is related to the proper time $\tau$ by the relation
\begin{equation}
t=\frac ca \sinh \left( \frac{a \tau}c \right) .
\label{eq:9}
\end{equation}

We also assume $\bx (0)=0$, $\dot{\bx}(0)=0$, and take into account that the two atoms are moving in a direction orthogonal to their distance, so that their distance does not change with time. We thus obtain the polarization and R\"{o}ntgen components of the electric and magnetic fields for the uniformly accelerating dipole, evaluated at the position of the other dipole,

\begin{widetext}
\begin{equation}\label{eq:10}
E_i^{(pol)}(\br , t) = - \left( \frac{1}{\rho ^3} \hat{T}_{ij} \mu_j(t_r) + \frac{1}{c \rho ^2} \hat{T}_{ij} \dot{\mu}_j(t_r) + \frac{1}{c^3\rho } \hat{S}_{ij} \ddot{\mu}_j(t_r) \right) ,
\end{equation}

\begin{eqnarray}\label{eq:11}
E_i^{(Roe)}(\br , t) = &-& \left( \frac{1}{c^2\rho ^2} \dot x_i (t_r) \hat{\rho}_j \dot{\mu}_j(t_r) +  \frac{1}{c^2\rho ^2}  \ddot x_i (t_r) \hat{\rho}_j {\mu}_j(t_r) \right. \nonumber \\
&+& \left. \frac{1}{c^3\rho }  \dot x_i (t_r) \hat{\rho}_j \ddot{\mu}_j(t_r) + \frac{1}{c^3\rho }  \dot{\ddot x}_i (t_r) \hat{\rho}_j {\mu}_j(t_r)+2 \frac{1}{c^3\rho }  \ddot x_i (t_r) \hat{\rho}_j \dot{\mu}_j(t_r) \right) ,
\end{eqnarray}

\begin{equation}\label{eq:12}
B_i^{(pol)}(\br, t) = -\frac{\epsilon _{ikj}}{c \rho ^2} \hat{\rho }_k \dot\mu _j(t_r) - \frac{\epsilon _{ikj}}{c^2 \rho} \hat{\rho }_k \ddot\mu _j  (t_r),
\end{equation}

\begin{eqnarray}\label{eq:13}
B_i^{(Roe)}(\br, t) = &-&  \frac{1}{c \rho^2} \hat T_{ij} \epsilon _{jkl} \left( \frac 1\rho \mu_k(t_r) \dot x_l(t_r) + \frac 1c \mu_k(t_r) \ddot x_l(t_r) + \frac 1c \dot \mu_k(t_r) \dot x_l(t_r) \right) \nonumber \\
&-& \frac{1}{c^3 \rho} \hat S_{ij} \epsilon _{jkl} \left( \mu_k(t_r) \dot{\ddot x}_l(t_r) + 2 \dot \mu_k(t_r) \ddot x_l(t_r) + \ddot \mu_k(t_r) \dot x_l(t_r) \right),
\end{eqnarray}
\end{widetext}
where $t_r=t-r/c$ is the retarded time, $\bm{\rho} (t) = \br - \bx (t)$ and $\epsilon _{ilk}$ is the totally antisymmetric symbol.
We have also defined the following tensors
\begin{eqnarray}
\hat{T}_{ij} &\equiv& \delta _{ij} - 3 \hat{\rho }_i \hat{\rho }_j , \label{eq:14a} \\
\hat{S}_{ij} &\equiv& \delta _{ij} - \hat{\rho }_i \hat{\rho }_j. \label{eq:14b}
\end{eqnarray}

In order to obtain the dispersion interaction energy for the two accelerating atoms, we need some considerations about the retarded time and the distance between the atoms to be used in the expressions for the fields. The effective interaction distance $\rho (t_r)$, introduced at the beginning of this Section, is the distance traveled by a light signal from one atom to the other one. For atoms at rest, it coincides with the interatomic distance $\rho$, while in the case of atoms moving at a constant velocity $v$, it is easy to show that
$\rho (t_r)= \gamma \rho$.  In our case the atoms are in an accelerated motion: this makes evident that we should expect an explicit time-dependence of the interaction distance because $\gamma = (1-v^2/c^2)^{-1/2}$ depends on time. By assuming that at $t=0$ the atoms are at rest and have a uniform acceleration $a$, using \eqref{eq:8} and simple geometrical considerations, it is possible to show that
\begin{equation}
\rho(t_r) = \rho + c\left (t - \frac {c\arctan \left (\frac {at} {c} \right)} {a} \right),
\label{eq:15}
\end{equation}
showing that indeed $\rho (t_r)$ depends on time and, as expected, it grows with time.

\section{\label{sec:3}The van der Waals interaction energy for the accelerating atoms}

We now evaluate the interaction energy between the fluctuating atomic dipoles in accelerated motion. We assume a nonrelativistic motion for the atoms; because their acceleration is given, this assumption limits the timescale of validity of our results, as we shall discuss in more detail in the next Section. The potential energy will be evaluated in the co-moving frame of the accelerating atoms. All relevant physical quantities will be however expressed in terms of quantities measured in the laboratory reference frame and thus directly measurable; this makes our approach different with respect to results in the literature concerning with radiative processes in accelerated frames (such as Lamb shift, atom-wall interactions, etc), which are in terms of physical quantities measured in the co-moving frame \cite{Passante98,RS09,ZY10,MP10}.

In our approach, each Fourier component of vacuum field fluctuations induces an oscillating dipole in the atoms, that in the laboratory frame is of the form (in the $\bk$ space)
\begin{equation}
\pmb \mu^{A(B),ind} (\bk ,\lambda , t) = \pmb \mu^{A(B),ind} (\bk ,\lambda ) \cos{(\omega t)}
\label{eq:16}
\end{equation}
with $\omega = c k$. Using \eqref{eq:1b}, the van der Waals interaction energy can be expressed as
\begin{equation}
\Delta \tilde{E} =  \sum _{\bk , \lambda} \sum _{\bk ', \lambda '} \tilde{\mu}^{B, ind}_i (\bk , \lambda ) \mu^{A, ind}_j(\bk ', \lambda ')
\tilde{V_{ij}}^\prime (\bR, t).
\label{eq:16a}
\end{equation}

We stress that in \eqref{eq:16a} the dipole moment of atom A is in the laboratory frame while that of atom B is still in the co-moving frame. We shall now transform the latter in the laboratory frame, in order to express the energy shift in terms of quantities in this frame only. Under a Lorentz transformation, the dipole moment transforms as a length; thus, in our case of atoms moving along the $x$ direction we have
\begin{equation}
\tilde{\pmb \mu} = \gamma \mu_x \hat{\imath}  + \mu_y \hat{\jmath}  +\mu_z \hat{k},
\label{eq:17}
\end{equation}
that shows that only the $x$ component is different in the two reference frames.

Using the relation \eqref{eq:2} between the induced dipole moment and the fluctuating vacuum field, we get
\begin{eqnarray}
\Delta \tilde{E} &=&  \sum _{\bk , \lambda} \sum _{\bk ', \lambda '}  \alpha (A,k') \alpha (B,k)
\nonumber \\
&\times& \! \! E_i(\bk ', \lambda '; \bR_A) E_j(\bk , \lambda; \bR_B) \tilde{V}_{ij}(\bR , t).
\label{eq:18}
\end{eqnarray}

The Fourier components of the electric field operator in \eqref{eq:18} are in the laboratory frame, because they come from relation \eqref{eq:2} with the induced dipole moment in the laboratory system. The tensor $\tilde{V}_{ij}(\bR , t)$ in \eqref{eq:18} differs from the tensor $\tilde{V_{ij}}^\prime (\bR, t)$ in \eqref{eq:16a} because the $\gamma$ factor in \eqref{eq:17} has been included in it, that is
\begin{eqnarray}
\tilde V_{xj}(\bR, t) &=& \gamma \tilde V_{xj}^\prime (\bR, t), \nonumber \\
\tilde V_{yj}(\bR, t) &=& \tilde V_{yj}^\prime (\bR, t), \hspace{12pt}
\tilde V_{zj}(\bR, t) = \tilde V_{zj}^\prime (\bR, t).
\label{eq:19}
\end{eqnarray}

In \eqref{eq:18} a factor $2$ should be added, taking into account that we should also consider an equal interaction energy obtained by exchanging the role of the two atoms, given by the interaction of the field emitted by atom B with atom A. We shall include this factor $2$ in the expression of the potential tensor $\tilde{V}_{ij}(\bR , t)$ given in the following of this Section.
We now take the vacuum expectation value of \eqref{eq:18}, taking into account that the electric field operators are in the laboratory frame. Thus we have
\begin{eqnarray}
&\ &\langle 0| E_i(\bk^\prime, \lambda^\prime; \bR_A) E_j(\bk, \lambda; \bR_B) |0 \rangle \nonumber \\
&=& \frac{2\pi \hbar c k}{V}\, {\hat{e}_i}(\bk, \lambda) {\hat{e}_j}^*(\bk, \lambda)\, e^{-i \bk\cdot (\bR_B -\bR_A)} \delta_{\bk \bk^\prime} \delta_{\lambda \lambda^\prime}.
\label{eq:20}
\end{eqnarray}

In the continuum limit, $V \to \infty$, $\sum_\bk \to V/(2\pi )^3 \int k^2 dk d\Omega$; performing polarization sum and angular integration,
\begin{equation}
\sum_\lambda \hat e_i(\bk, \lambda) \hat e_j^*(\bk, \lambda) = \delta _{ij} - \hat k_i \hat k_j ,
\label{eq:20a}
\end{equation}
\begin{eqnarray}
&\frac{1}{4\pi}& \int \left( \delta _{ij} - \hat k_i \hat k_j \right)\, e^{\pm i \bk \cdot \bR} \, d\Omega = \left( \delta _{ij} - \hat R_i \hat R_j \right) \frac{\sin(kR)}{kR} \nonumber \\
&+&\left( \delta _{ij} - 3\hat R_i \hat R_j \right) \left( \frac{\cos(kR)}{k^2 R^2}-\frac{\sin(kR)}{k^3 R^3} \right),
\label{eq:21}
\end{eqnarray}
we obtain
\begin{widetext}
\begin{equation}
\langle \Delta \tilde{E}\rangle  =  2  \frac{\hbar c}{\pi }\int \left\{ \hat S_{ij}\, \frac{\sin(kR)}{kR}+\hat T_{ij} \,\left( \frac{\cos(kR)}{k^2 R^2}-\frac{\sin(kR)}{k^3 R^3} \right)\right\}\, \tilde V_{ij}(\bR, t) \,k^3 \,dk.
\label{eq:22}
\end{equation}
\end{widetext}
In the approximation of a nonrelativistic motion, we have $\dot x(t)= a t$, $\ddot x(t)=a$ and $\dot{\ddot x}(t)=0$. Using these expressions in (\ref{eq:10}-\ref{eq:13}), we obtain the electric and magnetic fields generated by the uniformly accelerating dipole in the laboratory frame. In order to obtain the expression of the tensor $\tilde V_{ij}(\bR, t)$ in (\ref{eq:22}), we need the electric field in the co-moving frame. Thus we Lorentz-transform the fields according the well-known relations
\begin{eqnarray}
\tilde E_x &=& E_x \nonumber \\
\tilde E_y &=& \gamma (E_y - \beta B_z) \nonumber \\
\tilde E_z &=& \gamma  (E_z +\beta  B_y).
\label{eq:23}
\end{eqnarray}
\cite{Jackson98}. Using these transformations, the potential tensor $\tilde V_{ij}(\bR, t)$ in (\ref{eq:22}) is obtained as
\begin{widetext}
\begin{align}
\tilde V_{1j}(\bR ,t) = &- \frac{2\gamma (t)}R \left\{ \frac{\hat T_{1j}}R \left[ -\frac 1R \mathcal{A}(R,t) + \frac \omega c \mathcal{B}(R,t) \right] +
\hat S_{1j} \frac{\omega^2}{c^2}\mathcal{A}(R,t) + \hat R_j \frac a{c^2} \left[ \left( \frac 1R + \frac {\omega^2 t}c \right)\mathcal{A}(R,t)
\right. \right. \notag \\
& + \left. \left. \left( \frac {\omega t}R + \frac {2\omega}c \right) \mathcal{B}(R,t) \right] \right\} ,
\label{eq:24a}
\end{align}
\begin{align}
\tilde V_{2j}(\bR ,t) = &-\frac {2\gamma (t)}{R} \left\{ \frac{\hat T_{2j}}R \left[ -\frac 1R \mathcal{A}(R,t) +\frac \omega c \mathcal{B}(R,t) \right]
+ \hat S_{2j} \frac {\omega ^2}{c^2} \mathcal{A}(R,t)
- \frac {\beta (t)}c \left[ \hat R_{l} \varepsilon_{3lj} \omega \left( \frac \omega c \mathcal{A}(R,t) + \frac 1R \mathcal{B}(R,t) \right) \right. \right.
\notag \\
&+ \left. \left. \hat T_{3l} \varepsilon _{lj1} \frac aR \left( -\left( \frac 1c + \frac tR \right) \mathcal{A}(R,t) +
\frac {t \omega}c \mathcal{B}(R,t) \right)
+ \hat S_{3l} \varepsilon _{lj1} \frac {a\omega}{c^2} \left[ \omega t \mathcal{A}(R,t) + 2 \mathcal{B}(R,t) \right] \right] \right\} ,
\label{eq:24b}
\end{align}
\begin{align}
\tilde V_{3j}(\bR ,t) = &-\frac {2\gamma (t)}{R} \left\{ \frac{\hat T_{3j}}R \left[ -\frac 1R \mathcal{A}(R,t) +\frac \omega c \mathcal{B}(R,t) \right]
+ \hat S_{3j} \frac {\omega ^2}{c^2} \mathcal{A}(R,t)
+ \frac {\beta (t)}c \left[ \hat R_{l} \varepsilon_{2lj} \omega \left( \frac \omega c \mathcal{A}(R,t) + \frac 1R \mathcal{B}(R,t) \right) \right. \right.
\notag \\
&+ \left. \left. \hat T_{2l} \varepsilon _{lj1} \frac aR \left( -\left( \frac 1c + \frac tR \right) \mathcal{A}(R,t) +
\frac {t \omega}c \mathcal{B}(R,t) \right)
+ \hat S_{2l} \varepsilon _{lj1} \frac {a\omega}{c^2} \left[ \omega t \mathcal{A}(R,t) + 2 \mathcal{B}(R,t) \right] \right] \right\} ,
\label{eq:24c}
\end{align}
\end{widetext}
where $\beta (t)=v(t)/c$, $\gamma (t) = (1- \beta^2(t))^{-1/2}$. We have used (\ref{eq:14a}) and (\ref{eq:14b}) with $\bR$ in place of $\bm{\rho}$, and defined the functions
\begin{eqnarray}
\mathcal{A}(R,t) &=& \cos(\omega t)\, \cos\left[\omega \left( t-\frac{R}{c} \right)\right] , \label{eq:25a}\\
\mathcal{B}(R,t) &=& \cos(\omega t)\, \sin\left[\omega \left( t-\frac{R}{c} \right)\right] . \label{eq:25b}
\end{eqnarray}

Some considerations about the time-dependence of the potential tensor $\tilde V_{ij}(\bR, t)$ are now necessary. In the case of atoms at rest in the laboratory system, discussed in \cite{PT93}, the potential tensor is calculated, for each mode $(\bk , \lambda )$, after a time average on an oscillation period $2\pi /\omega$ of the dipoles. In that case, this is equivalent of taking a time-average of the quantities $\mathcal{A}(R,t)$ and $\mathcal{B}(R,t)$ in (\ref{eq:25a}) and (\ref{eq:25b}), respectively. In our case of accelerating atoms, extra time dependence is contained in the factors $\beta (t)$ and $\gamma (t)$ in Equations (\ref{eq:24a}-\ref{eq:24c}). We now take the time-average of $\tilde V_{ij}(\bR, t')$ on a time $t$ much larger than $\omega^{-1}$ (that is we take $\omega t \gg 1$ for a given $\omega$) and keep the leading term in $t$ only, which gives the main contribution to the time average. We thus consider the quantity
\begin{equation}
\langle \tilde V_{ij}(\bR , t) \rangle = \frac 1t \int_0^t V_{ij}(\bR , t') dt'.
\label{eq:26}
\end{equation}
We take a nonrelativistic approximation; then
\begin{equation}
\beta (t) \simeq  \frac{at}{c}; \, \,
\gamma (t) \simeq 1+ \frac{a^2 t^2}{2 c^2}
\label{eq:27}
\end{equation}
and keep only terms up to the second order in $at/c$.
In order to evaluate (\ref{eq:26}) we need to calculate integrals of $\mathcal{A}(R,t')$ and $\mathcal{B}(R,t')$ and integrals of these functions multiplied by $t'$ or ${t'}^2$, keeping only leading terms in $t$. After lengthy but straightforward algebraic calculations, we finally obtain
\begin{align}
& \langle \tilde V_{ij}(\bR , t) \rangle = \left( 1+ \frac{a^2 t^2}{6 c^2}\right)
\frac 1{R^3} \left\{\hat T_{ij}\left[ \cos(kR) \right. \right.\notag \\
&+ \left. \left. kR\sin(kR)\right] \hat S_{ij} \,k^2R^2\sin (kR) \right\} + Z_{ij} ,
\label{eq:28}
\end{align}
where $\bR = (0,0,R)$ is along the $z$ axis, and
$\hat T_{ij}= \mbox{diag}(1,1,-2)$ and $\hat S_{ij}= \mbox{diag}(1,1,0)$ are diagonal $3$x$3$ matrices. The $3$x$3$ matrix $Z_{ij}$ is defined below.
Substituting (\ref{eq:28}) into (\ref{eq:22}), we obtain the van der Waals interaction energy shift of the two accelerating atoms
\begin{widetext}
\begin{equation}
\langle \Delta \tilde{E}\rangle  =  \left( 1+ \frac{a^2 t^2}{6 c^2}\right)\, \Delta E^r+ 2  \frac{\hbar c}{\pi }\int \left\{ \hat S_{ij}\, \frac{\sin(kR)}{kR}+\hat T_{ij} \,\left( \frac{\cos(kR)}{k^2 R^2}-\frac{\sin(kR)}{k^3 R^3} \right)\right\}\, Z_{ij} \,k^3 \,dk
\label{eq:29}
\end{equation}
where
\begin{align}
\Delta E^r &= - \frac{\hbar c}{\pi R^3} \int _0^\infty \alpha (A; k)\alpha(B; k)
\left[ kR \, \sin(2kR) + 2 \, \cos(2kR) - 5\frac{\sin(2kR)}{kR}-6\frac{\cos(2kR)}{k^2R^2}+3\frac{\sin(2kR)}{k^3R^3}\right] k^3\, dk \notag \\
 &= - \frac{\hbar c}{\pi R^2} \int _0^\infty \alpha (A; iu)\alpha(B; iu) \left[ 1 + \frac{2}{uR} + \frac{5}{u^2R^2}+\frac{6}{u^3R^3}+\frac{3}{u^4R^4}\right] u^4 e^{-2uR}\, du
\label{eq:30}
\end{align}
\end{widetext}
is the well-known van der Waals potential energy for two atoms at rest \cite{CP48,PT93}.
In (\ref{eq:29}), $t$ is the observation time and $\langle \Delta \tilde{E}\rangle$ is the interaction energy averaged between times $0$ and $t$, as it follows from our averaging in (\ref{eq:26}); however, for sake of simplicity, we shall call it as the interaction energy at time $t$.
Eq. (\ref{eq:29}) clearly shows that one effect of the uniform acceleration of the atoms is a correction to the potential proportional to $a^2t^2/c^2$ and a new term (that with the $k$ integral), that we are now going to evaluate explicitly. We will show that this new term gives also a change of the $R$-dependence of the van der Waals potential energy when the two atoms are subjected to a uniform acceleration. From (\ref{eq:29}) and taking into account that $\hat T_{ij}$ and $\hat S_{ij}$ are diagonal matrices, we notice that only diagonal elements of the matrix $Z_{ij}$ appearing in (\ref{eq:28}) are relevant. Their values are
\begin{equation}
Z_{11} = 0
\label{eq:31a}
\end{equation}
\begin{eqnarray}
Z_{22} &=& \hat T_{33}\left[\frac{a^2t}{2c^3 R^2} \cos(kR)+\frac{a^2t^2}{3c^2 R^3}\cos(kR) \right.
\nonumber \\
&+ &\left. \frac{a^2t^2}{3c^2 R^2}k\sin(kR)\right]
\label{eq:31b}
\end{eqnarray}
\begin{eqnarray}
Z_{33} &=& \hat T_{22}\left[\frac{a^2t}{2c^3 R^2} \cos(kR)+\frac{a^2t^2}{3c^2 R^3}\cos(kR) \right. \nonumber \\
&+& \left. \frac{a^2t^2}{3c^2 R^2}k\sin(kR)\right]  \nonumber \\
&+& \hat S_{22}\left[\frac{a^2t}{c^3 R} k \sin(kR)-\frac{a^2t^2}{3c^2 R}k^2\cos(kR)\right].
\label{eq:31c}
\end{eqnarray}

Substitution of (\ref{eq:31a}-\ref{eq:31c}) into (\ref{eq:29}), finally yields
\begin{widetext}
\begin{align}
\langle \Delta \tilde{E}\rangle  = &\Delta E^r+\frac{a^2 t}{2 c^3}\frac{\hbar c}{\pi R^3} \int _0^\infty \alpha (A; iu)\alpha(B; iu) \, \left( 3 + \frac{4}{uR} +\frac{2}{u^2R^2}\right)u^2\,e^{-2uR}\, du\,+\notag \\
 +&\frac{a^2t^2}{6c^2}\frac{\hbar c}{\pi R^2} \int _0^\infty \alpha (A; iu)\alpha(B; iu) \, \left( -1+\frac{4}{uR}+\frac{8}{u^2R^2}+\frac{8}{u^3R^3}+\frac{4}{u^4R^4}\right) u^4\,e^{-2uR}\,du
 \label{eq:32}
\end{align}
\end{widetext}

The result given by (\ref{eq:32}) shows two terms correcting the van der Waals potential energy due to the atomic uniform acceleration: both are proportional to the square of the acceleration, and they explicitly depend on time as $t$ and $t^2$, within our approximations. Because the potential for inertial atoms $\Delta E^r$ is negative (attractive interaction), Equation (\ref{eq:32}) shows that the effect of the acceleration is to reduce the interaction energy, and this reduction grows with time. This is consistent with the fact that the ``effective interaction distance'' $\rho(t_r)$ in (\ref{eq:15}) grows as time goes on, yielding a decrease of the interaction energy between the accelerating atoms. However, as we shall discuss in more detail in Section \ref{sec:4}, these corrections cannot turn the potential in a repulsive one, i.e. a positive value of $\langle \Delta \tilde{E}\rangle$, at least within our approximations.

In order to discuss in more detail the effect of the acceleration, we can consider two limiting cases of the van der Waals dispersion energy, the so-called {\it near zone} and {\it far zone}. The near zone is when the interatomic distance $R$ is smaller than a main transition wavelength of the atoms; the far zone (Casimir-polder regime) is for larger distances.

In the near zone, the interaction energy $\Delta E^r$ for atoms at rest is as $R^{-6}$. In this zone, we can approximate $uR \ll 1$ in (\ref{eq:32}), obtaining
\begin{align}
\langle \Delta \tilde{E}\rangle &\simeq - \left(1 -\frac{4a^2t^2}{9c^2}\right)\frac{3\hbar c}{2\,\pi R^6} \int _0^\infty \alpha (A; iu)\alpha(B; iu) \,du
\notag \\
&+ \frac{a^2t\, \hbar}{\pi \,c^2 R^5} \int _0^\infty \alpha (A; iu)\alpha(B; iu) \,du.
\label{eq:33}
\end{align}

In the far zone we can approximate the atomic dynamical polarizabilities to their static value $\alpha^{A,B} (0)$, obtaining
\begin{widetext}
\begin{align}
\langle \Delta \tilde{E}\rangle  = &\Delta E^r  -  \alpha^A (0)\alpha^B(0)\, \frac{\hbar c}{\pi R^3}\int _0^\infty \left\{ \frac{a^2 t}{2 c^3 k}
\left[ 3\sin(2kR) + 4\frac{\cos(2kR)}{kR} -2 \frac{\sin(2kR)}{k^2R^2} \right] \right. \notag \\
&+ \left.  \frac{a^2t^2}{6c^2} \left[ kR \, \sin(2kR) -2 \, \cos(2kR) +3\frac{\sin(2kr)}{kR}+2\frac{\cos(2kR)}{k^2R^2}-\frac{\sin(2kR)}{k^3R^3}\right] \right\} k^3\, dk \, ,
\label{eq:34}
\end{align}
\end{widetext}
where in this case (far zone) the dispersion energy $\Delta E^r$ behaves as $R^{-7}$.
Performing the $k$ integrals, we finally get
\begin{align}
\langle \Delta \tilde{E}\rangle  \simeq &-\frac{\hbar c}{\pi}\frac{\alpha^A(0) \alpha^B(0)}{R^7} \left( \frac{23}{4} - \frac{7}{24} \frac{a^2 t^2}{c^2} \right)
\notag \\
&+ \frac{11\hbar\, a^2 t}{8\pi\,c^2}\frac{\alpha^A(0) \alpha^B(0)}{R^6}.
\label{eq:35}
\end{align}

These results clearly show the two new main features of the van der Waals interaction energy for accelerating atoms: a change of the dependence on the distance and an explicit time-dependence. In fact, from Equation (\ref{eq:35}) we can see that in the far zone an effect of the acceleration is to add a new (time-dependent) term behaving as $R^{-6}$, which has a longer range than the usual $R^{-7}$ van der Waals energy in the Casimir-Polder regime for atoms at rest. A $R^{-6}$ term in the atom-atom dispersion energy is known to occur when the interaction is calculated for atoms at rest at finite temperature \cite{Barton01}, and this indicates the deep connection between our results and the Unruh effect. The near-zone result (\ref{eq:33}) also shows corrections giving an explicit time-dependence of the interaction energy proportional to the acceleration squared, and a new term proportional to acceleration and time, and decreasing as $R^{-5}$. This new term has thus a slower decrease with the interatomic distance compared with the $R^{-6}$ behavior for inertial atoms. Also, the explicit time dependence as $a^2t^2/c^2$ in the first line of both (\ref{eq:33}) and (\ref{eq:35}), for the near and far zone respectively, gives corrections to the interaction that grows with time and may become significant even for a reasonable value of the acceleration. In fact, it is possible to find time intervals such that, from one side the nonrelativistic approximation is still valid ($a^2t^2/c^2 \ll 1$), and on the other side the corrective term, although relatively small, is not negligible.
For example, if $a^2t^2/c^2 \simeq 0.2$, we can still consider reasonable our approximation of a nonrelativistic motion of the atoms, and the correction to the van der Waals interaction energy from (\ref{eq:33}) and (\ref{eq:35}) is around ten percent in the near zone and one percent in the far zone. These changes are small, but not negligible.
Because only the product of acceleration and time is relevant for this correction to the dispersion energy due to the acceleration (and not the absolute value of the acceleration, as in the correction to the Lamb shift or the atom-wall interaction energy \cite{AM94,Passante98,RS09,ZY10}), this should be  achievable even with reasonable accelerations, provided a sufficiently long time is taken.
Also the corrections as $R^{-6}$ and $R^{-5}$ in the second lines of (\ref{eq:33}) and (\ref{eq:35}), respectively, give a change to the van der Waals interaction of a few percent, using the same value of the acceleration considered above and an interatomic distance $R$ such that $aR/c^2 \sim 0.1$, for which our use of a locally inertial system is valid (see also the discussion at the end of next Section).
All this suggests a new possibility for detecting the Unruh effect, or in general effects related to accelerated motion in quantum electrodynamics, without the extremely high accelerations necessary in the case of other quantum-electrodynamical effects recently discussed in the literature \cite{CHM08,Passante98,RS09,ZY10}.

\section{\label{sec:4}Concluding remarks}

We have considered the van der Waals interaction energy between two ground-state atoms (or polarizable bodies) moving in the vacuum with the same uniform acceleration. The acceleration is assumed orthogonal to the separation between the atoms, so that their distance is constant. In order to calculate their interaction energy, we have used a method based on the interaction between the instantaneous atomic dipole moments, which are induced and correlated by the zero-point fluctuations of the quantum electromagnetic field. We have shown there are two main effects of the acceleration: an explicit time-dependence of the the interatomic interaction and a qualitative change of its dependence from the interatomic distance, making the interaction of longer range. In particular, in the near zone a new term as $R^{-5}$ adds to the usual $R^{-6}$ behavior, while in the far zone a $R^{-6}$ term adds to the usual $R^{-7}$ van der Waals energy in the Casimir-Polder regime.

We now discuss some consequences of our results as well as the limits of our approximations.

Our result (\ref{eq:32}) for the van der Waals dispersion interaction energy for two uniformly accelerating atoms, and approximated in (\ref{eq:33}) and (\ref{eq:35}) for the near- and far-zone respectively, clearly shows how the accelerated motion of the atoms affects their interaction energy. Main effects are the time-dependence of the dispersion energy and a change of its distance dependence, depending on the acceleration squared, yielding a longer-range interatomic potential. This in an important point showing that the effect of the accelerated motion is not only a correction to the strength of the potential energy, but also a qualitative change of its properties. This also suggests, in perspective, the intriguing possibility of detecting signatures of the Unruh effect in interacting atomic systems, in particular when their properties, even at the macroscopic level, may critically depend on the form of the interaction among the atoms.
The time dependence of the interaction in (\ref{eq:32}) is related to the effective interaction distance given by (\ref{eq:15}), which grows with time for the accelerated atoms, making larger the ``effective distance'' traveled by the virtual photons exchanged between the atoms, as time goes on. A similar effect is not present in cases previously considered for the Lamb shift of an accelerated hydrogen atom \cite{AM94,Passante98} or the atom-surface Casimir-Polder interaction for an atom accelerating parallel to an infinite conducting plate \cite{RS09}: in these cases, the field fluctuations perceived by the atom are time-independent and the atom-surface ``effective distance'' is constant, and thus a time-dependence is not expected and the corrections depend on the absolute value of the acceleration only. We have also shown that taking appropriate values of the product of acceleration and time, the relative change of the van der Waals interaction, with parameters such that all our approximations are valid, can be in the range 1-10 percent, and thus not negligible.

In our model, we have neglected the possibility that the atoms are excited due to their acceleration. It is known that accelerated atoms have a finite probability of being spontaneously excited \cite{AM94,ZYL06,BC08}. In principle, this could add another source of change of the distance dependence of the dispersion interaction between the atoms, because this interaction behaves differently if one or both atoms are excited \cite{PT93a}. This probability, however, behaves as $1/(e^{2\pi c\omega_0/a}-1)$,
$\omega_0$ being a main atomic transition frequency \cite{AM94,ZYL06,BC08}.  It is thus
very small (exponentially) when $a \ll c\omega_0$. Taking a typical value for $\omega_0 \sim 10^{15} \, \text{s}^{-1}$, we expect that this contribution be negligible for $a \ll 10^{23} \, {\rm m/s^2}$. This is an extremely high acceleration, and our results show that we can obtain a significant change of the van der Walls energy for much smaller accelerations (making negligible the excitation probability, which decreases exponentially with decreasing accelerations), provided we consider a sufficiently long time (see discussion above). Thus atomic excitation induced by acceleration can be neglected in our case. Moreover, the contribution of the atomic excitation to the interatomic potential energy is a higher-order effect. In fact, the van der Waals interaction is a fourth-order effect, both for ground- and excited-state atoms \cite{PT93a}. Because the atomic excitation probability due to acceleration is a second-order effect, its contribution to the van der Waals interaction starts from sixth-order in the atom-field interaction.

Finally, we wish to make some considerations about the sign of the interaction energy of the accelerated atoms, which determines the attractive or repulsive character of the electric van der Waals force between two ground-state atoms (for atoms at rest it is always attractive). Equations (\ref{eq:33}) and (\ref{eq:35}) show that the accelerated motion reduces the potential energy between the atoms; this reduction grows with time, in agreement with the increasing effective interaction distance given by (\ref{eq:15}). One interesting question is to investigate whether the terms related to the acceleration in (\ref{eq:33}) and (\ref{eq:35}) can turn the van der Waals force to a repulsive character, that is making the interaction energy positive. In the near zone, analyzing Eq. (\ref{eq:33}) we see that the $R^{-6}$ term changes sign when $at/c$ is of the order of unity, but this is not compatible with our nonrelativistic approximation. On the other hand, the new (positive) $R^{-5}$ term becomes comparable with the usual (negative) $R^{-6}$ term for a distance between the atoms
$R \sim c^3/(a^2t)$ and, due to our nonrelativistic approximation $at/c \ll 1$, this would require $R \gg c^2/a$. This situation, however, would require a different treatment of our problem, by quantizing the field in a curved space-time; in fact, our use of a locally inertial system for the accelerated atoms is valid only when the dimension of the system is much less than $c^2/a$ \cite{Fulling89,MTW73,MF08}. An interatomic distance larger than $c^2/a$ cannot thus be considered by adopting the locally inertial frame we have used. Similar considerations hold for the far-zone potential energy in (\ref{eq:35}), too. We can then conclude that, within our approximation scheme, the attractive character of the van der Waals interaction is preserved also for the accelerated atoms. However, our results show that the van der Waals interaction between the two atoms is significantly affected by their acceleration, as shown by equations \eqref{eq:29}, \eqref{eq:33} and \eqref{eq:35}. In particular, as already mentioned, the time-dependence of the interaction energy could allow to detect the accelerated motion without necessity of the extremely high accelerations necessary in the case of other quantum electrodynamics effects recently proposed in the literature.

\begin{acknowledgments}
The authors wish to thank G. Compagno, J. Marino and L. Rizzuto for interesting discussions on subjects related to this work.
Financial support by the Julian Schwinger Foundation, by Ministero dell'Istruzione, dell'Universit\`{a} e della
Ricerca, by Comitato Regionale di Ricerche Nucleari e di Struttura della Materia, and by the ESF Research Networking Program CASIMIR is
gratefully acknowledged.
\end{acknowledgments}


\begin{references}
\bibitem{CP48} H. B. G. Casimir, D. Polder, Phys. Rev.  {\bf 73}, 360 (1948).
\bibitem{DMRR11} D. A. R. Dalvit, P. W. Milonni, D. C. Roberts, and F. S. S. Rosa in \emph{Casimir Physics}, edited by D. Dalvit, P. Milonni, D. Roberts, F. Rosa, Springer, Berlin 2011, p. 1.
\bibitem{Milonni07} P.W. Milonni, Physica Scripta {\bf 76}, C167 (2007).
\bibitem{Buhmann12} S. Y. Buhmann, \emph{Dispersion Forces I}, Springer, Berlin 2012.
\bibitem{MVP10} R. Messina, R. Vasile, and R. Passante, Phys. Rev. A {\bf 82}, 062501 (2010).
\bibitem{BP99} M. A. Bevan and D. C. Prieve, Langmuir {\bf 15}, 7925 (1999).
\bibitem{BCMZS10} H. Bender, Ph. W. Courteille, C. Marzok, C. Zimmermann, and S Slama, Phys. Rev. Lett. {\bf 104}, 083201 (2010), and references therein.
\bibitem{BVCLB13} L. Beguin, A. Vernier, R. Chicireanu, T. Lahaye, and A. Browaeys, Phys. Rev. Lett. {\bf 110}, 263201 (2013).
\bibitem{OWAPSC07} J. M. Obrecht, R. J. Wild, M. Antezza, L. P. Pitaevskii, S. Stringari, and E. A. Cornell, Phys. Rev. Lett. {\bf 98}, 063201 (2007).
\bibitem{Unruh76} W. G. Unruh, Phys. Rev. D {\bf 14}, 870 (1976).
\bibitem{CHM08} L. C. B. Crispino, A. Higuchi, and G. E. A. Matsas, Rev. Mod. Phys. {\bf 80}, 787 (2008).
\bibitem{BL83} J. S. Bell and J. M. Leinaas, Nucl. Phys. B {\bf 212}, 131 (1983).
\bibitem{CT99} P. Chen and T. Tajima, Phys. Rev. Lett. {\bf 83}, 256 (1999).
\bibitem{SCH06} R. Sch\"{u}tzhold, G. Schaller, and D. Habs, Phys. Rev. Lett. {\bf 97}, 121302 (2006).
\bibitem{US12} W. G. Unruh and R. Sch\"{u}tzhold, Phys. Rev. D {\bf 86}, 064006 (2012).
\bibitem{BCCGORRSF10} F. Belgiorno, S. L. Cacciatori, M. Clerici, V. Gorini, G. Ortenzi, L. Rizzi, E. Rubino, V. G. Sala, and D. Faccio, Phys. Rev. Lett.
{\bf 105}, 203901 (2010).
\bibitem{AM95} J. Audretsch and R. M\"{u}ller, Phys. Rev. A {\bf 52}, 629 (1995).
\bibitem{Passante98} R. Passante, Phys. Rev. A {\bf 57} 1590 (1998).
\bibitem{RS09} L. Rizzuto and S. Spagnolo, Phys. Rev. A {\bf 79}, 062110 (2009).
\bibitem{ZY10}  Z. Zhu and H. Yu, Phys. Rev. A {\bf 82}, 042108 (2010).
\bibitem{ZYL06} Z. Zhu, H. Yu, and S. Lu, Phys. Rev. D {\bf 73}, 107501 (2006).
\bibitem{MP10} J. Marino and R. Passante, in: \emph{Quantum Field Theory under the Influence of External Conditions (QFEXT09)}, edited by K.A. Milton and M. Bordag, World Scientific, Singapore 2010, p. 328.
\bibitem{PT93} E. A. Power, T. Thirunamachandran, Phys. Rev. A {\bf 48}, 4761 (1993).
\bibitem{PPR03} R. Passante, F. Persico, L. Rizzuto, Phys. Lett. A {\bf 316}, 29 (2003).
\bibitem{CP97} M. Cirone and R. Passante, J. Phys. B {\bf 30}, 5579 (1997).
\bibitem{SPR06} S. Spagnolo, R. Passante, and L. Rizzuto, Phys. Rev. A {\bf 73}, 062117 (2006).
\bibitem{Salam09} A. Salam, J.Phys.: Conf. Ser. {\bf 161}, 012040 (2009).
\bibitem{PT01} E. A. Power, T. Thirunamachandran, Proc. R. Soc. Lond. A {\bf 457}, 2757 (2001).
\bibitem{Thirunamachandran06} T. Thirunamachandran, J. Phys. B {\bf 39}, S725 (2006).
\bibitem{Rindler91} W. Rindler, \emph{Introduction to Special Relativity}, Oxford University Press, Oxford 1991.
\bibitem{Jackson98} J. D. Jackson, \emph{Classical Electrodynamics}, Wiley 1998.
\bibitem{Barton01} G. Barton, Phys. Rev. A {\bf 64}, 032102 (2001).
\bibitem{Fulling89} S. A. Fulling, \emph{Aspects of Quantum Field Theory in Curved Space-Time}, Cambridge University Press, Cambridge 1989.
\bibitem{MTW73} C. W. Misner, K. S. Thorne, and J. A. Wheeler, \emph{Gravitation}, W.H. Freeman and Co, San Francisco 1973
\bibitem{MF08} J.W. Maluf and F. F. Faria, Ann. Phys. (Berlin) {\bf 17}, 326 (2008).
\bibitem{AM94} J. Audretsch and R. M\"{u}ller, Phys. Rev. A {\bf 50}, 1755 (1994).
\bibitem{BC08} G. Barton and A. Calogeracos, J. Phys. A {\bf 41}, 164030 (2008).
\bibitem{PT93a} E. A. Power and T. Thirunamachandran, Phys. Rev. A {\bf 47}, 2539 (1993).
\end{references}
\end{document}